\documentclass[aps,prb,reprint,superscriptaddress,showpacs,twocolumn]{revtex4}
\usepackage{units}
\usepackage{amsmath}
\usepackage{amssymb}
\usepackage{graphicx}
\usepackage{bm}
\usepackage{microtype,color}

\newcommand{\be}{\begin{equation}}
\newcommand{\ee}{\end{equation}}
\newcommand{\bea}{\begin{eqnarray}}
\newcommand{\eea}{\end{eqnarray}}

\begin{document}

\title{Probing Long-Range Intensity Correlations inside Disordered Photonic Nanostructures}

\author{Raktim Sarma}
\affiliation{\textls[-20]{Department of Applied Physics, Yale University, New Haven, CT, 06520, USA}}
\author{Alexey Yamilov}
\email{yamilov@mst.edu}
\affiliation{\textls[-20]{Department of Physics, Missouri University of Science and Technology, Rolla, Missouri 65409,USA}}
\author{Pauf Neupane}
\affiliation{\textls[-20]{Department of Physics, Missouri University of Science and Technology, Rolla, Missouri 65409,USA}}
\author{Boris Shapiro}
\affiliation{\textls[-20]{Department of Physics, Technion-Israel Institute of Technology, Haifa 32000, Israel}}
\author{Hui Cao}
\email{hui.cao@yale.edu}
\affiliation{\textls[-20]{Department of Applied Physics, Yale University, New Haven, CT, 06520, USA}}
%$\footnote{e-mail:hui.cao@yale.edu}$
\date{\today}
\begin{abstract}

We report direct observation of the development of long-range spatial intensity correlations and the growth of intensity fluctuations inside the random media. We fabricated quasi-two-dimensional disordered photonic structures and probed the light transport from a third dimension. Good agreements between experiment and theory are obtained. We were able to manipulate the long-range intensity correlations and intensity fluctuations inside the disordered waveguides by simply varying the waveguide geometry.

\end{abstract}


\maketitle

\section{Introduction}

Light propagation in disordered media has been a topic of intense studies for nearly three decades \cite{Sheng1, Feng3, Rossum1}. In analogy to electronic transport in disordered metals, fundamental issues related to diffusion and localization have been addressed \cite{Sheng2, Akkermanbook}. One interesting example is the long-range intensity correlations\cite{Cwilich}, which characterize the mesoscopic transport of both classical and quantum waves, and reflect the closeness to the Anderson localization threshold \cite{Pninichapter}. Experimentally correlations in time, space, frequency, angle, and polarization have been investigated, but most measurements are performed on transmitted or reflected light outside the random media \cite{Maret1,Sebbah1,Garcia2,Genack1,Genack2,Lagendijk1,Lagendijk2,Chabanov1,Muskens1}.
It would be interesting to probe correlations inside the random media and monitor how long-range correlations build up as light propagates through the random media. However, it is very difficult to probe transport inside three-dimensional (3D) random media. Only in the microwave experiment, a detector (antenna) has been inserted into the random media to measure the intensity inside\cite{Genack1}. Alternatively we design and fabricate quasi-2D disordered waveguides to probe light transport inside from the third dimension \cite{Dz}. This approach will allow us to directly measure the intensity correlations and fluctuations inside the random structures. Furthermore, we vary the degree of long-range intensity correlations by changing the waveguide geometry.

The intensity-intensity correlation function $C$ consists of three terms, a short-range $C_1$, a long-range $C_2$ and an infinite-range $C_3$. Intuitively interferences between waves scattered along independent paths give rise to $C_1$, one crossing of paths generates $C_2$, and two crossings cause $C_3$\cite{Feng1,Feng2}. The spatial correlation $C_1$ decays exponentially with increasing distance and vanishes beyond the transport mean free path $\ell$. $C_2$ also decays but much more slowly, while $C_3$ has a constant contribution. The long-range correlation leads to a fluctuation of total transmission $T_a \equiv \sum_{b} T_{ab}$, where $T_{ab}$ is the transmission from an incoming wave mode $a$ to an outgoing mode $b$. The magnitude of $C_2$ is on the order of $1/g$, and $C_3$ of $1/g^2$, where $g \equiv \sum_a T_a$ is the conductance \cite{Maret1,Maret2}. When $g \gg 1$, $C$ is dominated by $C_1$. To measure $C_2$, the spatial distance must exceed the transport mean free path so that $C_1$ dies out. Alternatively, $C_2$ can be measured by collecting all transmitted light using an integrating sphere. This method, however, cannot be adopted for the measurement of $C_2$ {\it inside} the sample. Instead, we integrate light intensity over the waveguide cross-section to average out the short-range fluctuation, and directly measure the long-range correlation inside the disordered {\it planar waveguide}. The conductance of the waveguide is $g =(\pi/2) N\ell/L$, where $N = 2W/(\lambda/n_e)$ is the number of propagating modes in the waveguide. $L$ is the waveguide length, $W$ is the waveguide width, $\lambda$ is the light wavelength in vacuum, and $n_e$ is the effective index of refraction of the random medium. Hence, by decreasing $W$, we are able to reduce $g$ and enhance the magnitude of $C_2$ without modifying the structural disorder.

%\bibitem{Qiu_neff} \cc{M.~Qiu, Appl. Phys. Lett. {\bf 81}, 1163 (2002).}

This paper is organized as follows. In section II, we describe the design and fabrication of 2D disordered waveguides as well as the optical measurement of intensity correlations inside the waveguide. Section III contains the calculation of long-range correlations inside the disordered waveguides and the formula for the physical quantities that are measured experimentally. Section IV presents the experimental data and comparison to the theory. Finally we conclude in Section V.

\section{2D disordered photonic structures}

The 2D disordered waveguides were fabricated in a silicon-on-insulator (SOI) wafer with a 220 nm silicon layer on top of a 3$\mu m$ buried oxide [Fig. 1]. The patterns were written by electron beam lithography and etched in an inductively-coupled-plasma (ICP) reactive-ion-etcher (RIE). Each waveguide contained a 2D random array of air holes that scattered light. The air hole diameters were 100 nm and the average (center-to-center) distance of adjacent holes was 390 nm. The waveguide walls were made of photonic crystals (triangle lattice of air holes, the lattice constant = 440 nm, the hole radius = 154 nm) that had complete 2D bandgap for in-plane confinement of light. However, light was scattered out of plane, and this leakage allowed us to observe light transport inside the disordered waveguide from the vertical direction.

\begin{figure}[htbp]
\centering
\includegraphics[width=1\linewidth]
{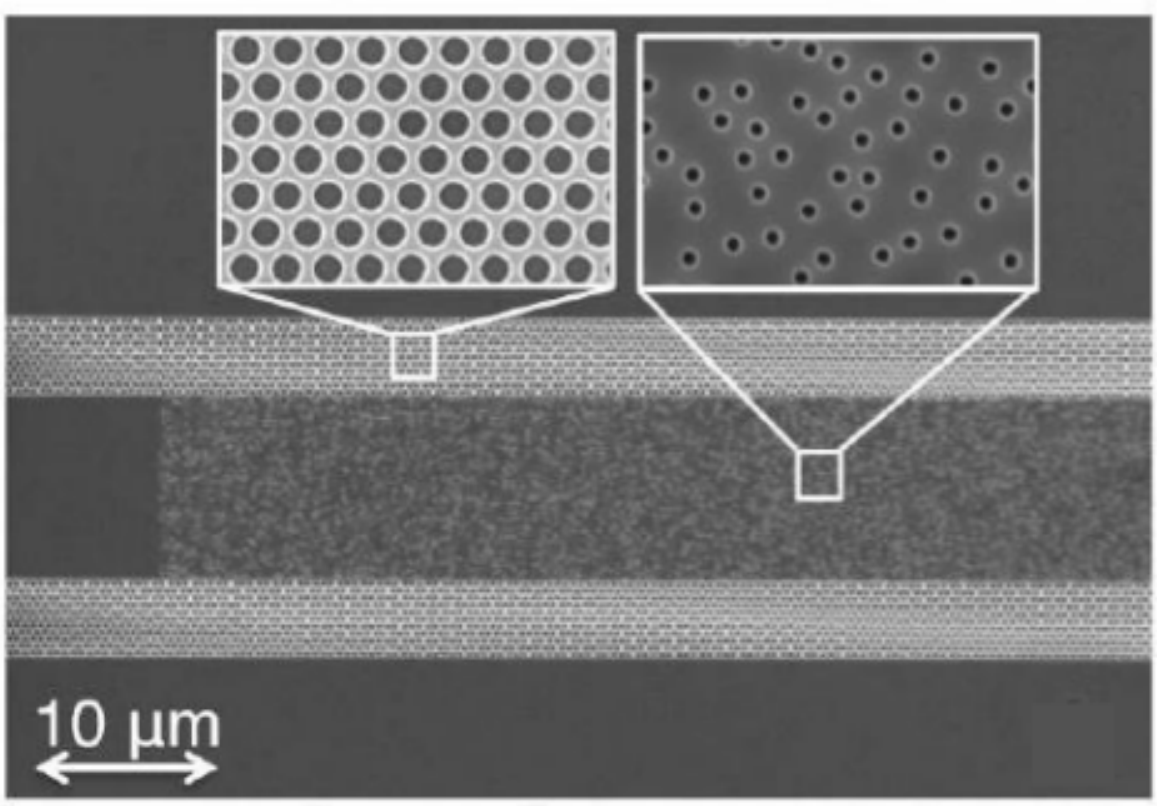}
\caption{Top-view scanning electron microscope (SEM) image of a quasi-2D disordered photonic waveguide. Light is injected from the left end of the empty waveguide and incident onto the random array of air holes. The waveguide wall is made of a triangle lattice of air holes which forms a 2D photonic bandgap to confine light inside the waveguide.}
\end{figure}

The monochromatic light from a tunable CW laser source (HP 8168F) was coupled by an objective lens of numerical aperture (NA) = 0.4 into the empty waveguide. To ensure efficient confinement inside the waveguide, the light was transverse-electric (TE) polarized (electric field in the plane of the waveguide). It was subsequently incident onto the random array of air holes inside the waveguide and underwent multiple scattering [Fig. 2(a)]. The near-field optical image of the spatial distribution of light intensity across the structure  was taken by collecting light scattered out of plane using a 50$\times$ objective lens (NA = 0.42) and recorded by an InGaAs camera (Xeva 1.7-320). The spatial resolution was limited by the NA of the objective lens, and estimated to be $\sim$ 2 $\mu m$. Figure 2(b) is a typical near-field image, which exhibits the short-range intensity fluctuations.

\begin{figure}[htbp]
\centering
\includegraphics[width=1\linewidth]
{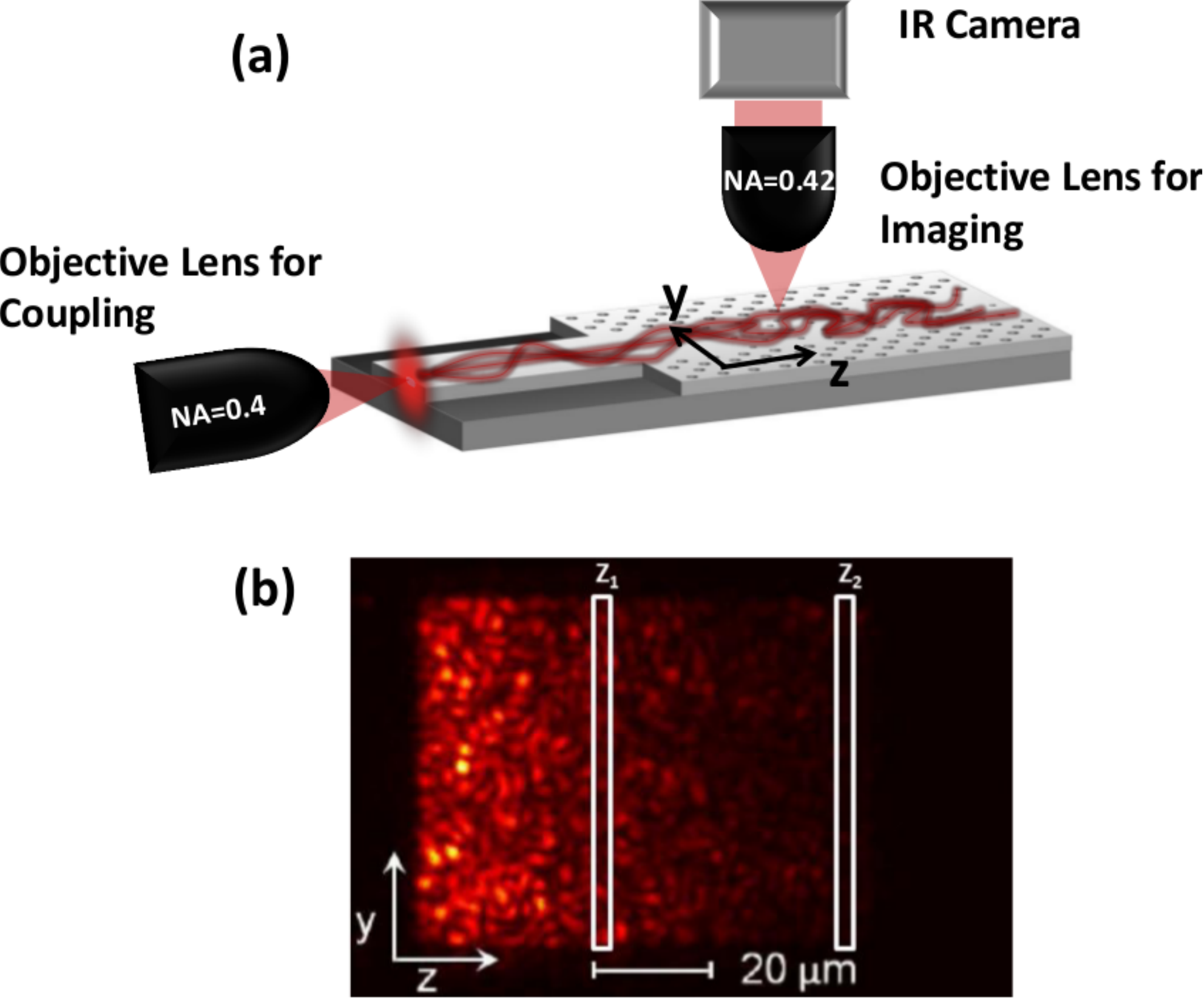}
\caption{(Color online)(a)A schematic of the optical measurement setup. One objective lens (NA = 0.4) couples the light from a tunable laser source to the waveguide, and another objective lens ($50 \times$, NA = 0.42) collects the light scattered by the air holes out of the waveguide plane and images onto a camera.  (b) A near-field optical image of the intensity of scattered light from the disordered waveguide. The wavelength of the probe light is 1510 nm. The intensity distribution exhibits short-range fluctuations. $z_1$ and $z_2$ represent the axial positions of two cross-sections inside the disordered waveguide. }
\end{figure}

From the near-field image [Fig. 2(b)], the 2D intensity distribution inside the waveguide $I(y,z)$ was extracted. Then $I(y,z)$ was integrated along the cross-section of the waveguide ($y$ direction) to give the variation along the waveguide axis ($z$ direction) $I(z)$. The spatial intensity correlations were then computed from $I(z)$ as:
\begin{equation}
\tilde{C}(z_1,z_2) = \frac{\langle I(z_1)I(z_2)\rangle}{\langle I(z_1)\rangle \langle I(z_2)\rangle}-1
\label{eq:C_definition}
\end{equation}
where $\langle..\rangle$ represents an ensemble average. $\tilde{C}(z_1,z_2)$ was measured for various combinations of $z_1$ and $z_2$ inside the disordered waveguides.  The ensemble averaging was done over ten random configurations of air holes and 25 input wavelengths equally spaced between 1500 nm and 1510 nm. The wavelength spacing was chosen to produce different intensity distributions. Further averaging was done by generating different intensity distributions by slightly moving the input coupling spot along the transverse direction $y$ . Nevertheless, since long range correlations depend on the size and shape of the input beam \cite{Shapiro2}, we ensured that the random array of air holes was illuminated uniformly along the $y$ direction, so that diffusion occurs only along the $z$ direction.

The relevant parameters for light propagation in the disordered waveguide are the transport mean free path $\ell$ and the diffusive dissipation length $\xi_a$. The transport mean free path $\ell$ depends on the size and density of the air holes.
The dissipation mostly comes from out-of-plane scattering as the silicon absorption at the probe wavelength is negligible.
As shown in our previous work\cite{Dz}, this vertical leakage of light can be treated similarly as absorption and described by the diffusive dissipation length $\xi_a = \sqrt{D \tau_a}$, where $\tau_a $ is the ballistic absorption time and $D$ is the diffusion coefficient
For the disordered waveguides in Fig. 1, we found $\xi_a = 30$ $\mu$m and $\ell = 2.2$ $\mu$m by fitting the measured $I(z)$ with the diffusion equation. A detailed description of this procedure is given in Ref.\cite{Dz}. The waveguide length is 80 $\mu$m, and the width varies from 10 $\mu$m to 60 $\mu$m. Thus the conductance $g$ is between 1.6 and 9.9.

\section{Theory of Long-Range Intensity Correlations}

Spatial intensity correlations defined by Eq.~(\ref{eq:C_definition}) involve intensities integrated over the cross-section of the waveguide. Such integration suppresses the contribution from the short-range correlation $C_1$ so that only $C_2$ and $C_3$ remain. At the output end of the disordered waveguide ($z_1=z_2=L$), these two contributions reduce to the normalized variance of total transmission and the normalized variance of conductance respectively \cite{Sebbah1,Akkermanbook}. $C_2$ and $C_3$ in lossy systems, such as those in our experiment,have been investigated before\cite{Brouwer1,Pninichapter}.
%\begin{widetext}
%\begin{eqnarray}
%C_2&=&\frac{1}{4 {\cal L} g} \left(2 {\cal L} + \coth{\cal L} - \frac{{\cal L}}{\sinh^2{\cal L}}\right)
%\label{eq:varTa}\\
%C_3&=&\left(\frac{\sinh {\cal L}}{{\cal L} g}\right)^2 \left(\frac{2 {\cal L}^2 - 9 {\cal L} \coth {\cal L} + 12}{16 \sinh^2{\cal L}} - %\frac{3 {\cal L}^2}{16 \sinh^4{\cal L}} + \frac{8 {\cal L} \coth {\cal L} - 11}{16 \sinh^2{\cal L}} + \frac{6 {\cal L}^2 - 3 {\cal L} %\coth {\cal L} - 3}{16 \sinh^4{\cal L}} + \frac{3 {\cal L}^2}{8 \sinh^6{\cal L}}\right)
%\label{eq:varg}
%\end{eqnarray}
%\end{widetext}
Although the expressions for $C_2$ and $C_3$ in Ref.\cite{Brouwer1,Pninichapter} have been derived for diffusive samples ($g>1$), it has been shown to also apply to the localized samples ($g\leq 1$) \cite{Yamilov2}. For the disordered waveguides in our experiment, $C_2$ is much larger than $C_3$. Thus we ignore $C_3$ and assume $\tilde{C} \simeq C_2$.

%Insert here a short paragraph by Alexey on the validity of using the 2D formula...

Next we obtain an expression for $C_2(z_1,z_2)$ which can be used to compare directly with the spatial correlation function defined in Eq.~(\ref{eq:C_definition}). Such expression has been derived using the Langevin approach in Refs.~\cite{Pnini1,Lagendijk1,Lisansk2,Lisansk1}. For a waveguide geometry we get,
\begin{equation}
C_2(z_1,z_2)=\frac{2}{gL}\frac{\displaystyle\int_0^L \frac{\partial K(z_1,z^\prime)}{\partial z^\prime} \frac{\partial K(z_2,z^\prime)}{\partial z^\prime} \langle I(z^\prime)\rangle^2 dz^\prime}{\langle I(z_1)\rangle\langle I(z_2)\rangle},
\label{eq:C2_expression}
\end{equation}
where $K(z,z^\prime)$ is the solution of
\begin{equation}
\frac{\partial^2 K(z,z^\prime)}{\partial z^2} -\frac{K(z,z^\prime)}{\xi_a^2}=-\delta(z-z^\prime)
\label{eq:kernel_equation}
\end{equation}
with boundary conditions $K(0,z^\prime)=K(L,z^\prime)=0$. This boundary condition neglects surface effects\cite{Lisansk1} that are important for $0<z\leq\ell$, $L-\ell\leq z<L$. This assumption is reasonable in our case, since $\ell\ll L$. Solution to Eq.~(\ref{eq:kernel_equation}) is
\begin{equation}
K(z,z^\prime)=\frac{\sinh\zeta_<\sinh({\cal L}-\zeta_>)}{\xi_a^{-1}\sinh{\cal L}},
\label{eq:K_solution}
\end{equation}
where ${\cal L}=L/\xi_a$, $\zeta_<=\min[z,z^\prime]/\xi_a$ and $\zeta_>=\max[z,z^\prime]/\xi_a$. In the same approximation $\langle I(z)\rangle\propto \sinh({\cal L}-\zeta)/\sinh{\cal L}$. Substituting this expression as well as Eq.~(\ref{eq:K_solution}) into Eq.~(\ref{eq:C2_expression}) we get $C_2(z_1,z_2)$. The final expression is cumbersome in presence of loss, so we only list several limiting cases.

Case 1: Vanishing loss. In this case we take the limit $\xi_a\rightarrow\infty$ and get
\begin{equation}
C_2(z_1,z_2)=\frac{2z_1}{3g L^2}\frac{(2L-z_1)(L-z_1)+(L-z_2)^2}{L-z_1},
\label{eq:C2_passive}
\end{equation}
which reduces to a well known result $C_2(L,L)=2/3g$ at the output end.

Case 2: $z_2=L$. This corresponds to correlating the intensity at the output surface with an intensity inside random medium. We get
\begin{widetext}
\begin{equation}
C_2(z_1,L)=
\frac{-8\zeta_1 +  4 \zeta_1 \cosh 2{\cal L} + 3(\sinh 2{\cal L}-\sinh 2\zeta_1) -
   3\sinh 2({\cal L}-\zeta_1) + 4({\cal L}-\zeta_1)\cosh{\cal L}\ {\rm csch}({\cal L}-\zeta_1)\sinh\zeta_1}
{16\ g\ {\cal L}\ \sinh^2{\cal L}}.
\label{eq:C2(z1,L)}
\end{equation}
\end{widetext}
where, $\zeta_1=z_1/\xi_a$. In lossless random medium the above expression reduces to $C_2(z_1,L)=2(2L-z_1)z_1/(g\ L^2)$, in agreement with the expression in Ref.~\cite{Pnini1,Kaveh1}.

Case 3: $z_1=z_2\equiv z$. Under this condition we obtain the normalized variance of the cross-section integrated intensity inside the waveguide,
\begin{widetext}
\begin{eqnarray}
C_2(z,z)=
\left[
4\zeta\cosh 2{\cal L} +
   5\sinh 2{\cal L} - \sinh 2({\cal L}-2\zeta) +
{\rm csch}^2({\cal L}-\zeta) (-4({\cal L}-\zeta) + \sinh 4({\cal L}-\zeta))\sinh^2\zeta
\right.\nonumber\\
\left.-4(2\zeta +\sinh 2({\cal L}-\zeta)+\sinh 2\zeta)
\right]/\left[16\ g\ {\cal L}\ \sinh^2{\cal L}\right].
\label{eq:C2(z,z)}
\end{eqnarray}
\end{widetext}
In the limit $z=L$ this quantity reduces to the normalized variance of the total transmission. In lossless medium Eq.~(\ref{eq:C2(z,z)}) reduces to a compact expression $C_2(z,z)=(2z/g L)(1-2z/3L)$. We note that this function takes the maximum value $(9/8) C_2(L,L)$ at $z=3L/4$, for any $L$.

In the following section we will compare the above theoretical predictions to the experimental data.

\section{Experimental results and comparison to theory}

Figure 3 shows the measured $\tilde{C}(z_1,z_2)$ for a disordered waveguide of  $L$ = 80 $\mu$m, $W$ = 60 $\mu$m, $\xi_a$ = 30 $\mu$m, $\ell$ = 2.2 $\mu$m. $z_1$ is varied between 0 and $L$ while $z_2$ is fixed at $L$ or $L/2$.  As the distance between $z_1$ and $z_2$ increases, $\tilde{C}(z_1,z_2)$ decays gradually. Even when the distance becomes much larger than the transport mean free path, the intensity correlation does not vanish. The correlation builds up further into the sample. As shown in the inset of Fig. 3, for a fixed distance $ \Delta z = z_2-z_1=10$ $\mu$m, $\tilde{C}$ grows as $z_2$ moves from $L/4$ to $L$. The experimentally observed long-range correlations inside the random system agree well to the theoretical predictions represented by the solid lines in Fig. 3.

\begin{figure}[htbp]
\centering
\includegraphics[width=0.8\linewidth]
{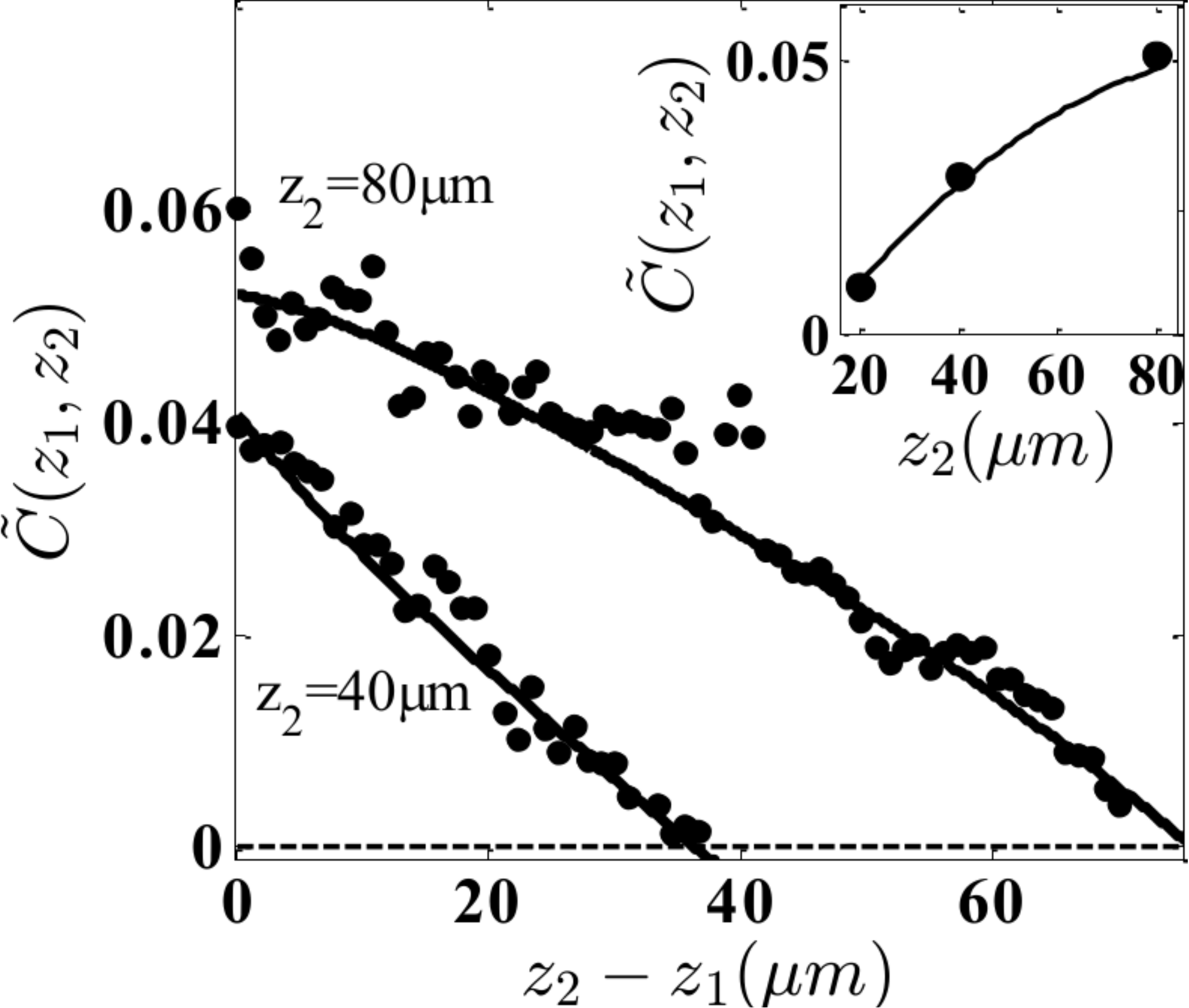}
\caption{Long-range intensity correlation $\tilde{C}(z_1,z_2)$ in a disordered waveguide of $L$ = 80 $\mu$m, $W$ = 60 $\mu$m, $\xi_a$ = 30 $\mu$m, $\ell$ = 2.2 $\mu$m. $z_1$ is varied between 0 and $L$ while $z_2$ is fixed at $L$ or $L/2$. Solid circles are experimental data and solid lines represent the theoretical predictions of Eqs.~(\ref{eq:C2_expression},\ref{eq:C2(z1,L)}). The dashed line corresponds to the background taken outside the waveguide. The inset shows $\tilde{C}(z_1,z_2)$ for $\Delta z = z_2-z_1$ = 10 $\mu$m and $z_2=L,L/2,L/4$. Solid circles are experimental data and solid line represents the theoretical prediction of Eqs.~(\ref{eq:C2_expression}).  For a fixed $\Delta z$, $\tilde{C}(z_1,z_2)$ increases when moving deeper into the sample.}
\end{figure}

Next we varied the width $W$ of the waveguide while keeping the length $L$ and the degree of disorder the same. Figure 4 compares $\tilde{C}(z_1,z_2)$ for two disordered waveguides of length $L$ =80 $\mu$m and $W$ = 10 $\mu$m, 60 $\mu$m. $z_1$ is moved from 0 and $L$ while $z_2$ is set at $L$.  The localization length $\xi$, reduces from 788 $\mu$m for $W$ = 60 $\mu$m to 131 $\mu$m for $W$ = 10 $\mu$m.  Hence, the former is in the diffusion regime $(\ell \ll L \ll \xi)$, while the latter approaches the localization regime $(L >  \xi)$. The conductance $g$, which is proportional to $W$, drops by a factor of 6 from 9.85 to 1.64. The probability for two scattering paths crossing, which scales as $1/g$, is thus enhanced by a factor of 6.  This leads to a six-fold increase of the long-range intensity correlation, as observed experimentally and confirmed theoretically. We note that the enhancement of long-range correlations, as a result of enhanced localization effect, is caused purely by the change of waveguide geometry with no modification of the scattering strength.

\begin{figure}[htbp]
\centering
\includegraphics[width=0.8\linewidth]
{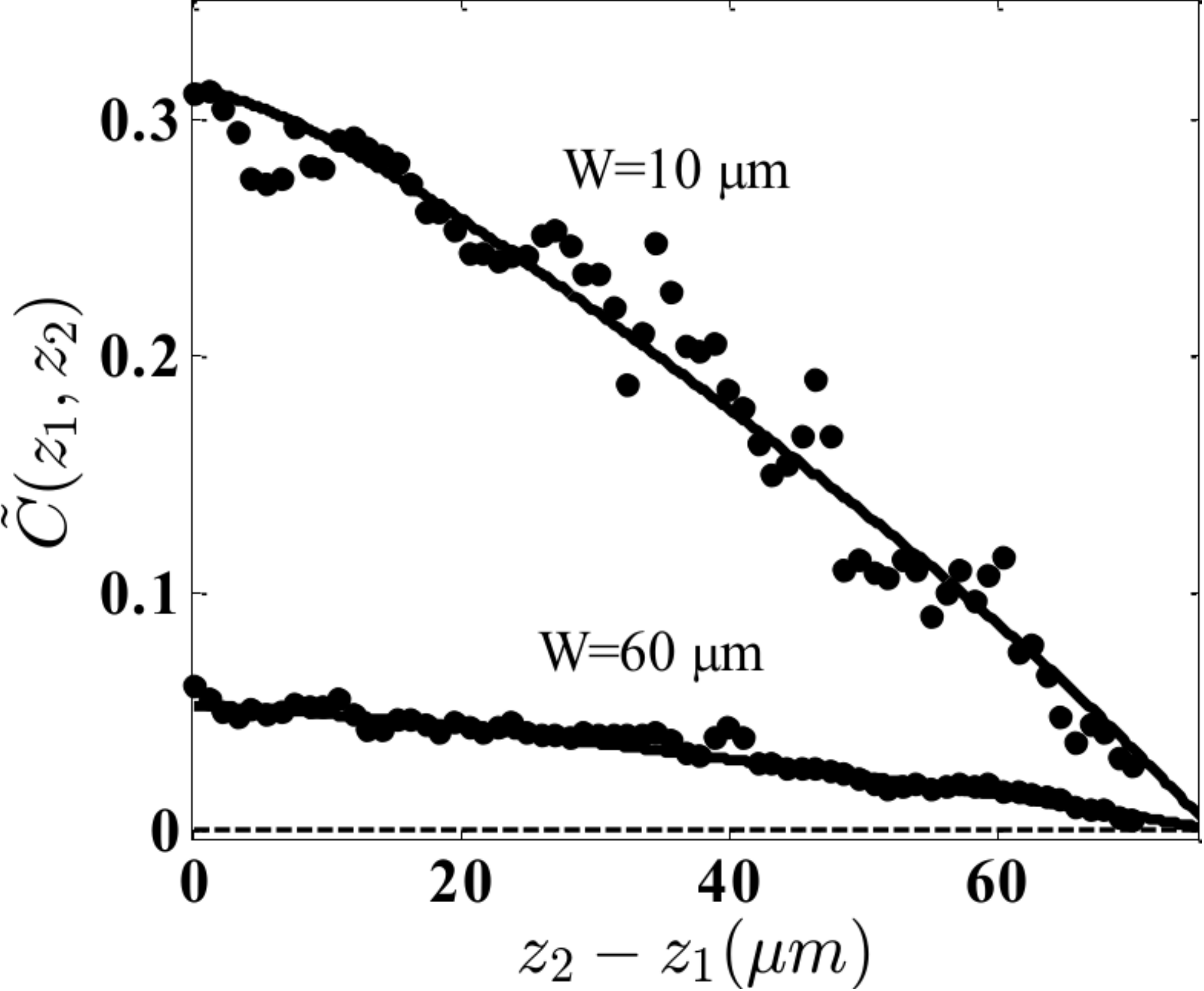}
\caption{Long-range intensity correlation $\tilde{C}(z_1,z_2)$ for two waveguides with the same length $L=80\mu m $ and the degree of disorder $(k\ell=26)$ but different widths $W $=60 $\mu$m and $W$ = 10 $\mu$m. $z_1$ is moved from 0 to $L$ and $z_2$ is set at $L$. Solid circles are experimental data and solid lines represent the theoretical predictions of Eqs.~(\ref{eq:C2(z1,L)}). The dashed line corresponds to the background taken outside the waveguide. The six-times reduction of the waveguide width results in a six-fold increase in the magnitude of intensity correlations.}
\end{figure}

Finally, we measured the variance of the cross-section integrated intensity $I(z)$ inside the disordered waveguides. As mentioned above, the normalized variance, $var[I(z)]/\langle I(z) \rangle^2 = \tilde{C}(z_1=z, z_2=z)$, becomes equal to the normalized variance of total transmission when $z=L$. Figure 5 shows the measured variance inside two disordered waveguides of width $W$ = 10 $\mu$m, 60 $\mu$m. The other parameters are the same as in Fig. 4. $z$ is changed from 0 to $L$. The fluctuation of $I(z)$ grows when moving deeper into the random system. In a narrower waveguide, the fluctuation is larger due to stronger localization effect  (smaller conductance).

\begin{figure}[htbp]
\centering
\includegraphics[width=0.8\linewidth]
{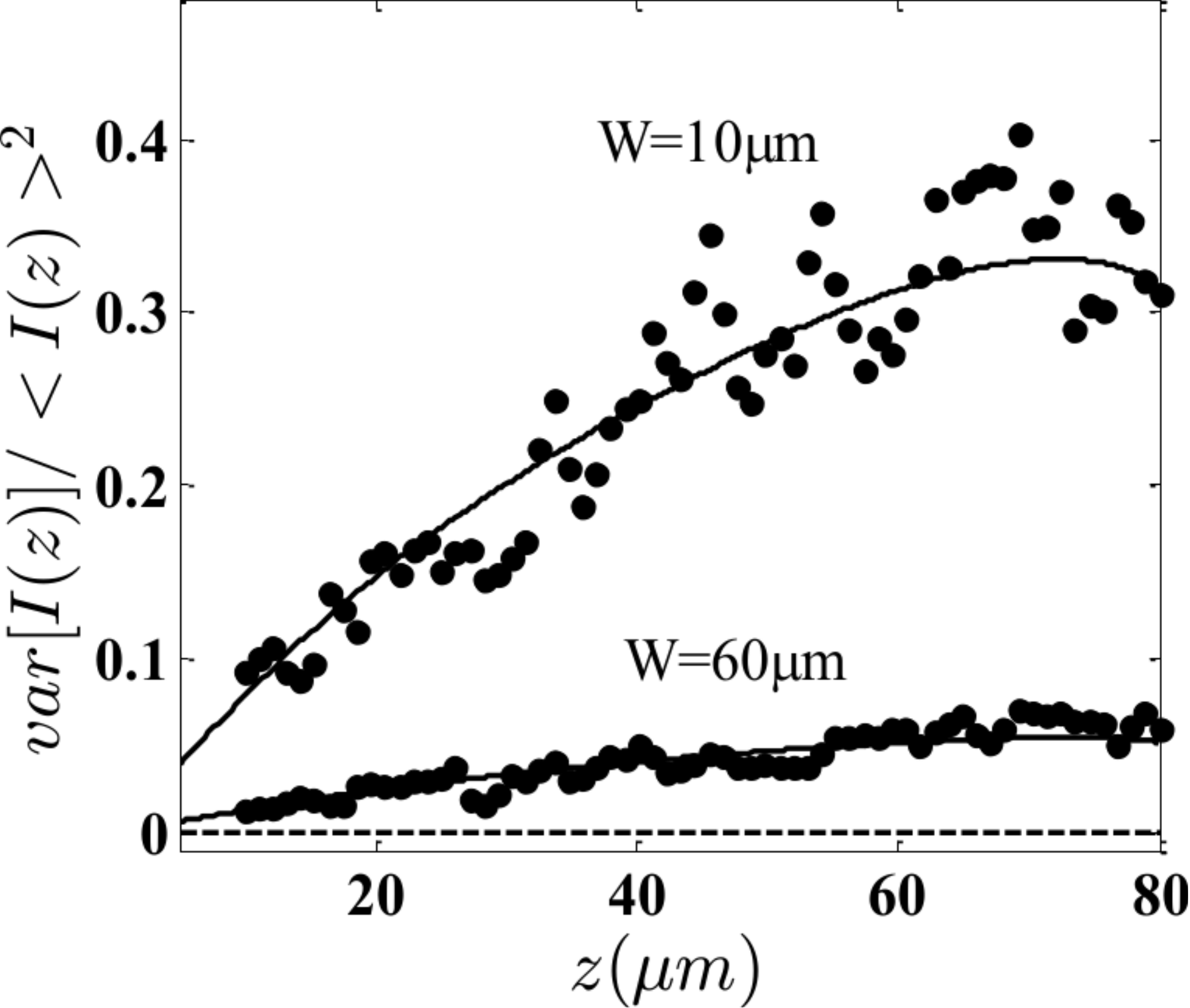}
\caption{Normalized variance of the cross-section integrated intensity $I(z)$, $var[I(z)]/\langle I(z) \rangle^2$, for two waveguides with the same length $L$ = 80 $\mu$m and degree of disorder $(k\ell=26)$ but different widths $W$ = 60 $\mu$m and $W$ = 10 $\mu$m. $z$ is changed from 0 to $L$. The solid circles are experimental data and solid lines represent the theoretical predictions of Eqs.~(\ref{eq:C2(z,z)}). The dashed line corresponds to the background taken outside the waveguide.The six-times reduction of the waveguide width results in a six-fold increase in the magnitude of intensity fluctuations. }
\end{figure}

\section{Conclusion}

In summary, we directly measured the long-range spatial intensity correlations inside the quasi-two-dimensional disordered waveguides. Light scattered out of the waveguide plane allowed us to probe the internal transport from the third dimension. The long-range intensity correlations gradually build up as light propagates through the random system. The fluctuations of cross-section integrated intensity also grow with the depth into the disordered waveguide. Good agreements between experiment and theory are obtained. By reducing the waveguide width, we are able to enhance the long-range intensity correlations and the intensity fluctuations, without modifying the degree of disorder. This provides a new approach for manipulation of long-range spatial correlations of light intensity inside random media.

%may have potential applications to present efforts in enhancing the capacity of wireless communication.

\section{Acknowledgment}

We acknowledge Douglas Stone and Arthur Goetschy for valuable discussions. We also thank Michael Rooks for useful suggestions on sample fabrication. This work was supported by the National Science Foundation under grants nos. DMR-1205307, DMR-1205223, ECCS-1128542 and ECCS-1068642. Facilities used were supported by YINQE and NSF MRSEC Grant No. DMR-1119826.


\begin{thebibliography}{99}

\bibitem{Sheng1}{\it Scattering and Localization of Classical Waves in Random Media},edited by P. Sheng (World Scientific, Singapore,1990).
\bibitem{Feng3} R. Berkovits and S. Feng, {\it Phys. Rep.} {\bf 238}, 135 (1994).
\bibitem{Rossum1} van Rossum, M. C. and Nieuwenhuizen, T. M., {\it Rev. Mod. Phys.} {\bf 71}, 313 (1999).
\bibitem{Sheng2} Ping Sheng, {\it Introduction to Wave Scattering, Localization, and Mesoscopic Phenomena}(Academic,Boston,1995).
\bibitem{Akkermanbook}E. Akkermans and G. Montambaux,{\it Mesoscopic Physics of Electrons and Photons}(Cambridge University Press, Cambridge,2007).
%\bibitem{Cwilich} \cc{M.~J.~Stephen and G.~Cwilich, Phys. Rev. Lett. {\bf 59}, 285 (1987).}
\bibitem{Cwilich} M.~J.~Stephen and G.~Cwilich, {\it Phys. Rev. Lett.} {\bf 59}, 285 (1987).
\bibitem{Pninichapter}R. Pnini,{\it Correlation of Speckle in Random Media, Proceedings of the International Physics School on Waves and Imaging through Complex Media}, 391–412 (1999: Cargèse, France) edited by P Sebbah (Kluwer Academic Publishers, Dordrecht, 2001).

\bibitem{Genack1}A. Z. Genack et al., {\it Phys. Rev. Lett.} {\bf 65}, 2129 (1990).
\bibitem{Lagendijk2}M. P. van Albada et. al., {\it Phys. Rev. Lett.} {\bf 64}, 2787 (1990).
\bibitem{Lagendijk1}Johannes F. de Boer et. al., {\it Phys. Rev. B} {\bf 45}, 658 (1992).
\bibitem{Garcia2}N.Garcia et al., {\it Phys. Lett. A} {\bf 176}, 458 (1993).
\bibitem{Maret1}Frank Scheffold et al., {\it Phys. Rev. B} {\bf 56}, 10942 (1997).
\bibitem{Genack2}P.Sebbah et al.,{\it Phys. Rev. E} {\bf 62}, 7348 (2000).
\bibitem{Sebbah1}P.Sebbah et al., {\it Phys. Rev. Lett.} {\bf 88}, 123901 (2002).
\bibitem{Chabanov1}A.A. Chabanov et. al., {\it Phys. Rev. Lett.} {\bf 92}, 173901 (2004).
\bibitem{Muskens1}O.L. Muskens et al., {\it Phys. Rev. B} {\bf 84}, 035106 (2011).
\bibitem{Dz}A.Yamilov et al., {\it Phys. Rev. Lett.} {\bf 112}, 023904 (2014).
\bibitem{Feng1}S. Feng, C. Kane, P. A. Lee, and A. D. Stone, {\it Phys. Rev. Lett.} {\bf 61}, 834 (1988).
\bibitem{Feng2}S. Feng and P. A. Lee, {\it Science} {\bf 251}, 633 (1991).

\bibitem{Maret2} Frank Scheffold and Georg Maret, {\it Phys. Rev. Lett.} {\bf 81}, 5800 (1998).
\bibitem{Shapiro2}R Pnini and B.Shapiro,{\it Phys. Rev. B} {\bf 39}, 6986 (1989).

\bibitem{Brouwer1}P. W. Brouwer, {\it Phys. Rev. B} {\bf 57}, 10526 (1998).

\bibitem{Yamilov2}A.Yamilov and H.Cao, {\it Phys. Rev. E.} {\bf 70}, 037603 (2004).

\bibitem{Pnini1}R.Pnini and B.Shapiro, {\it Phys. Lett. A} {\bf 157}, 265 (1991).
\bibitem{Lisansk2}A. A. Lisyansky and D. Livdan,{\it Phys. Lett. A} {\bf 170}, 53 (1992).
\bibitem{Lisansk1}A. A. Lisyansky and D. Livdan, {\it Phys. Rev. B} {\bf 47}, 14157 (1993).

\bibitem{Kaveh1}Eugene Kogan and Moshe Kaveh, {\it Phys. Rev. B} {\bf 45}, 1049 (1992).

\end{thebibliography}
 \end{document}